\documentclass[aps,prb,twocolumn,superscriptaddress,showpacs]{revtex4}
\usepackage{graphicx}
\usepackage{calc}
\usepackage{bm}

\begin{document}

\title{Equilibration between edge states in the fractional quantum Hall effect regime at high imbalances}

\author{E.V.~Deviatov}
\email[Corresponding author. E-mail:~]{dev@issp.ac.ru}
 \affiliation{Institute of Solid State
Physics RAS, Chernogolovka, Moscow District 142432, Russia}

\author{A.A.~Kapustin}
\affiliation{Institute of Solid State Physics RAS, Chernogolovka,
Moscow District 142432, Russia}

\author{V.T.~Dolgopolov}
\affiliation{Institute of Solid State Physics RAS, Chernogolovka,
Moscow District 142432, Russia}

\author{A.~Lorke}
\affiliation{Laboratorium f\"ur Festk\"orperphysik, Universit\"at
Duisburg-Essen, Lotharstr. 1, D-47048 Duisburg, Germany}

\author{D.~Reuter}
\affiliation{Lehrstuhl f\"ur Angewandte Festk\"orperphysik,
Ruhr-Universit\"at Bochum, Universit\"atsstrasse 150, D-44780
Bochum, Germany}

\author{A.D.~Wieck}
\affiliation{Lehrstuhl f\"ur Angewandte Festk\"orperphysik,
Ruhr-Universit\"at Bochum, Universit\"atsstrasse 150, D-44780
Bochum, Germany}

\date{\today}

\begin{abstract}
We experimentally study equilibration between edge states,
co-propagating at the edge of the fractional quantum Hall liquid,
at high initial imbalances. We find an anomalous increase of the
conductance between the fractional edge states at the filling
factor $\nu=2/5$ in comparison with the expected one for the
model of independent edge states. We conclude that the model
of independent fractional edge states is not suitable to describe
the experimental situation at  $\nu=2/5$.
\end{abstract}


\maketitle

In the integer quantum Hall effect (IQHE) regime, edge
states~\cite{halperin} (ES) are arising at the sample edge at the
intersections of the Fermi level and Landau levels. Buttiker
proposed a formalism~\cite{buttiker}, that allows to calculate
different transport characteristics of the sample by regarding the
transport through ES.  This picture was firmly confirmed in experiments
with crossing gates (for a review see Ref.~\onlinecite{haug}) and
in the quasi-Corbino geometry~\cite{alida}.

From the beginning, the fractional quantum Hall effect (FQHE) was
understood as the many-body phenomenon~\cite{laughlin}. Strongly
interacting electron system forms a new ground
state~\cite{laughlin,haldane,obzor}, that, contains 
gapless excitation modes at the sharp sample edges - fractional
ES~\cite{macdonald,wenPRB41}. While decreasing the sharpness of
the edge potential profile, edge reconstruction
occurs~\cite{chamon} and the sample edge is a set of
incompressible (with constant fractional filling factor) and
compressible electron liquids~\cite{Beenakker}, like in the
integer case~\cite{shklovsky}. Fractional ES are arising at the
edges of the incompressible stripes~\cite{chamon}. For
the calculation of the transport along the fractional edge,
Buttiker formulas can easily be modified~\cite{Beenakker}. These
formulas were validated in experiments on the transport along the
sample edge~\cite{crossgates}.

From both the experimental and theoretical points of view,
transport investigations \emph{across} the sample edge  should be
important. Fractional ES can be regarded as the realization of the
one-dimensional strongly-correlated electron
liquid~\cite{wenPRB41}, as was confermed in experiments on
tunnelling into the fractional
edge~\cite{milliken,pellegr,chang,grayson}. Except of the
tunnelling, even the equilibration between the fractional ES in
extended uniform  junctions is a point of question. It was
shown~\cite{fracbutt} to be sensitive to the internal
structure~\cite{chamon} of the incompressible stripes.
Authors~\cite{zhulike} concluded   that  interaction between ES
can significantly affect on the maximum  conductance of the line
 junction. Moreover, even the possibility to describe the
interacting fractional ES at high imbalance in terms of the local
electrochemical potentials is still an open
question~\cite{ponomarenko}. The fractional ES can be regarded as
independent only for very smooth edge potential
profile~\cite{Beenakker}, e.g. at the electrostatically defined
edge~\cite{kouwen,cunning}. The inter-ES interaction, however,
cannot be neglected at  stronger edge potentials, e.g. at etched
mesa edges, where the reconstructed fractional
edge~\cite{chamon,fracbutt} is expected.

Thus, to experimentally study the inter-ES equilibration,
investigations at imbalances higher than the spectral gaps should
be performed at etched mesa edge. This is impossible in the usual
Hall-bar technics~\cite{kouwen,cunning,komiyama}, but can be
easily performed in the quasi-Corbino sample
geometry~\cite{alida,weiss,relax}. Also, in view of the
theoretical investigations~\cite{wenPRB41,zhulike}, inter-ES
interaction differs qualitatively for different fractional
fillings, so measurements at 5th fractions are important.

Here we experimentally study equilibration between the fractional
ES, co-propagating at the same sample edge, at high initial
imbalances.  We find an anomalous increase of the conductance
between the fractional ES at the filling factor $\nu=2/5$ in
comparison with the expected one for the model of independent ES.
We conclude that the model of independent fractional
ES~\cite{Beenakker} is not suitable to describe the experimental
situation at $\nu=2/5$.

Our samples are fabricated from  molecular beam epitaxial-grown
GaAs/AlGaAs heterostructure. It contains a 2DEG located 150~nm
below the surface. The mobility at 4K is 1.83 $\cdot
10^{6}$cm$^{2}$/Vs and the carrier density 8.49 $\cdot
10^{10}$cm$^{-2}$,  as was obtained from usual magnetoresistance
measurements. Also, magnetocapacitance measurements were performed
to characterize the electron system under the gates. We use both
these methods to check the contact resistances and the sample
homogeneity. The cooling procedure with slow sample cooling was
used to obtain the well-reproducible, stable, and homogeneous
sample states, as was tested for 4 samples. This guaranties the
reliability of the results, presented below.

An interplay between two ground states~\cite{obzor} (spin
polarized (SP) at $B=5.18$~T and spin unpolarized (SU) at
$B=4.68$~T) at $\nu=2/3$ is well developed in our samples,
permitting the measurements at different spin configurations of
the $\nu=2/3$ ground state.

\begin{figure}
\includegraphics*[width=0.65\columnwidth]{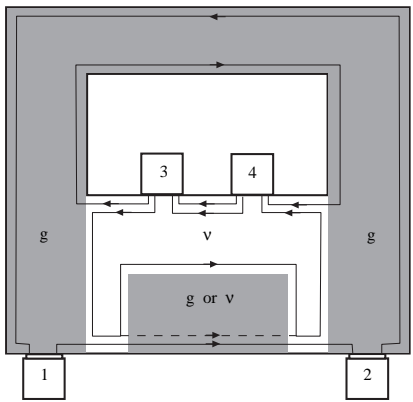}%
\caption{ Schematic diagram of the sample. Bold lines show etched
mesa edges. Bars with numbers denote the ohmic contacts. The light
gray areas represent the Schottky-gates. Arrows indicate the
direction of electron drift in the edge states. $\nu, g$ indicates
the filling factor in the corresponding region. \label{sample}}
\end{figure}

The quasi-Corbino sample geometry~\cite{alida,weiss,relax} is
modified for these measurements, see Fig.~\ref{sample}. Mesa of
the square form ($1.2\times 1.2$~mm$^2$) has a rectangular
($810\times 600 \mu\mbox{m}^2$) etched region inside it. Ohmic
contacts are made to both, the inner and the outer, mesa edges.
Two Schottky gates of the special form are placed on the top of
the crystal, allowing to diminish the electron concentration under
the gates. In the quantizing magnetic field at filling factor
$\nu$, the number of ES at the ungated mesa edges equals to $\nu$.
By depleting 2DEG under the main gate to the filling factor $g$,
some of ES (the number is $\nu-g$), are redirected to the other
mesa edge. Thus, ES from independent ohmic contacts run together
along the outer etched edge of the sample in the gate-gap region,
as depicted in Fig.~\ref{sample}. In the quantum Hall effect
regime (at integer or fractional $\nu, g$), a current between the
outer and inner ohmic contacts can only flow across the sample
edge in the gate-gap, because of the zero dissipative
conductivity. Auxiliary Schottky gate ($800\times 200
\mu\mbox{m}^2$) is placed into the gate-gap, allowing to control
the width of the interaction region. By depleting 2DEG to the same
filling factor $g$, it separates two groups of ES in the gate-gap
by the macroscopic distance (200$\mu$m). ES are running together
in two narrow (5$\mu$m) independent regions.  As a
result, there are two regimes of operation: (i) a negative bias is
applied to both gates, the interaction region is narrow ($2\times
5\mu$m$=10\mu$m); (ii) a negative bias is applied only to the main
gate, the auxiliary gate is grounded, the interaction region is wide
(810~$\mu$m).

 We study $I-V$ curves of the
gate-gap region in the 4-point configuration~\cite{alida}, that
allows to exclude any contact effects. We apply {\em dc} current
$I_{24}$ between the contacts no.2 and no.4 (grounded) and measure
{\em dc} voltage $V_{13}$ between the contacts no.1 and no.3, see
Fig.~\ref{sample}. For the case of full equilibration between ES
we can expect a linear $I-V$ with the equilibrium resistance that
can be determined in our geometry from Buttiker formulas for
integer~\cite{buttiker,alida}  or
fractional~\cite{Beenakker,fracbutt} fillings: $R_{eq}=h/e^2
\nu/g(\nu-g)$. We use a constant current mode to carefully study
linear regions of $I-V$'s (the resistances are in the range
1.5-28~$h/e^2$) and exclude contact resistances (0.5~kOhm), while
a constant voltage mode is more appropriate for strongly
non-linear $I-V$'s.  We checked that the interchange of the
current and voltage probes does not affect on the results,
presented here,  confirming they's reliability. The experiment is
performed in the dilution refrigerator with the base temperature
of 30~mK, equipped with the superconducting solenoid.

\begin{figure}
\includegraphics[width= 0.75\columnwidth]{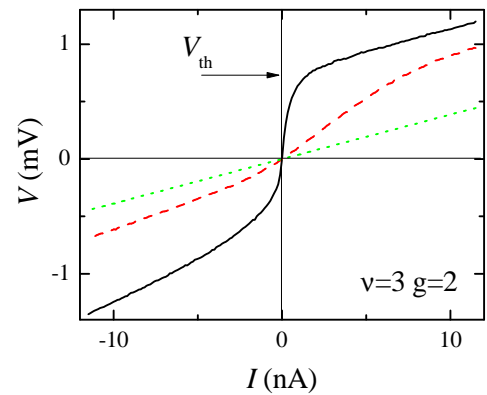}%
\caption{(Color online)  $I-V$ curves for integer filling factors
$\nu=3, g=2$ for narrow (10~$\mu$m, solid line) and wide
(800~$\mu$m, dashed line) interaction regions. Equilibrium curve
(with $R_{eq}=1.5 h/e^2$) is shown by dots. Magnetic field $B$
equals to 1.1~T. \label{IV32}}
\end{figure}

We start from the well-known situation~\cite{alida,relax} of
integer fillings. $I-V$ curves for the integer filling factors
$\nu=3, g=2$ are presented in Fig.~\ref{IV32}. The experimental
curve for the narrow gate-gap is strongly non-linear and
asymmetric. The positive branch is of threshold behavior, the
threshold value $V_{th}=0.73$~meV. After the threshold, the
positive branch goes with the constant slope, and is parallel to the
fully equilibrated ($R_{eq}=1.5 h/e^2$) theoretical line. The
slope of the negative branch of the $I-V$ trace is always higher
than the equilibrium $R_{eq}$. The experimental $I-V$ curve for
the wide gate-gap is also presented in the figure. The threshold
on the positive branch is still present, but is smaller and not so
well defined as in the previous case. The positive branch itself
is parallel to the equilibrium line at high currents. The negative
branch of the $I-V$ is still non-parallel to the equilibrium line.

\begin{figure}
\includegraphics[width= 0.8\columnwidth]{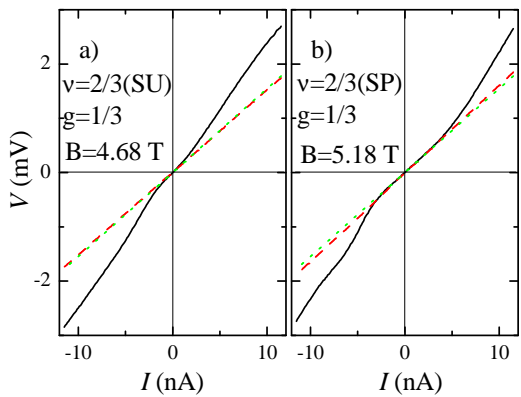}%
\caption{(Color online)   $I-V$ curves for fractional filling
factors $\nu=2/3, g=1/3$ for narrow (10~$\mu$m, solid line) and
wide (800~$\mu$m, dashed line) interaction regions, for two spin
configurations of $\nu=2/3$: (a) spin unpolarized (SU) state
($B=4.68$~T); (b) spin polarized (SP) state ($B=5.18$~T).
Equilibrium curve (with $R_{eq}=6 h/e^2$) is shown by dots.
\label{IV2313}}
\end{figure}

For the fractional filling factor combination $\nu=2/3, g=1/3$,
$I-V$ curves are shown in Fig.~\ref{IV2313} for both gate-gap
widths. $I-V$'s are presented in two panels, (a) and (b), for the
two different spin configurations of the $\nu=2/3$ ground state.
For the narrow gate-gap, $I-V$ curves are non-linear and close to
be symmetric. In contrast to the integer case, they have a linear
region near the zero without any threshold and two non-linear
branches. The central linear region with roughly  equilibrium
slope, is clearly defined in the spin-polarized (high-field)
state, but not so pronounced in the spin-unpolarized (low-field)
state. Non-linear branches disappear with increasing the
temperature up to $T=195$~mK, leaving the slope of the linear
central part to be unchanged. At the wide gate-gap the non-linear
branches are not present even at minimal temperature. $I-V$ curves
for both gate-gap widths coincide near the zero, the experimental
slope is not differ from the equilibrium $R_{eq}=6 h/e^2$ within
3$\%$, see Fig.~\ref{IV2313}.

\begin{figure}
\includegraphics[width= 0.75\columnwidth]{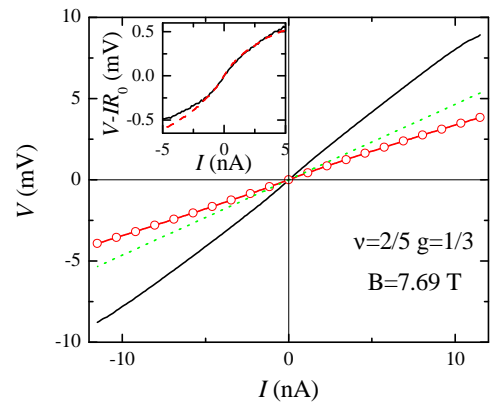}%
\caption{(Color online)  $I-V$ curves for fractional filling
factors $\nu=2/5, g=1/3$  for narrow (10~$\mu$m, solid line) and
wide (800~$\mu$m, line with open circles) interaction regions.
Equilibrium curve (with $R_{eq}=18 h/e^2$) is shown by dots. Inset
shows the wide-region curve (dash), scaled to the narrow-region
one (solid) in x-direction. The linear dependence $IR_0$ with
$R_0=28 h/e^2$ is subtracted to highlight the non-linear behavior.
Magnetic field $B$ equals to 7.69~T. \label{IV2513}}
\end{figure}

The most intriguing experimental result is the evolution of the
$I-V$ curve for the fractional filling factor combination
$\nu=2/5, g=1/3$, as shown in Fig.~\ref{IV2513}. The experimental
curve for the narrow gate-gap consists from two slightly
non-linear branches and is situated above the equilibrium line
($R_{eq}=18 h/e^2$). It is similar to the shown in
Fig.~\ref{IV2313} (a), but the central linear region is not
developed at all. Increasing the gate-gap width leads to the $I-V$
curve, which is situated \emph{below} the equilibrium one, see
Fig.~\ref{IV2513}, with 28\% lower resistance. This curve is still
non-linear and can be scaled to one for the narrow gate-gap by
dividing the current by factor $q=2.35$, see inset to
Fig.~\ref{IV2513}. Increasing the temperature up to $T=0.62$~K
results in linear $I-V$ traces with the slope, which is equal to
$5.1 h/e^2 \ll R_{eq}=18 h/e^2$ for both gate-gap widths. In other words, the scaling
coefficient $q$ is approaching to 1 with increasing the
temperature.

Thus, we have two most important experimental results: (i) for the
narrow gate-gap (10$\mu$m) $I-V$ curves are nonlinear for both the
integer and fractional fillings, but differ in the symmetry and
the zero-bias behavior (ii) for the wide gate-gap (800$\mu$m) the
slope of the fully equilibrated curve is significantly {\em
smaller} than the calculated equilibrium one for the fractional
filling factors $\nu=2/5, g=1/3$, in contrast to the integer case
and the simple fractional  $\nu=2/3, g=1/3$ fillings.

At real edge profiles, the edge of the sample is a set of
compressible and incompressible electron liquids in both the
IQHE~\cite{shklovsky} and FQHE~\cite{chamon,Beenakker} regimes.
Applying a voltage between the outer and inner ohmic contacts
leads to the electrochemical potential imbalance across the
incompressible stripe at the "injection" corner of the gate-gap
(the left one in Fig.~\ref{sample}). While going along the sample
edge in the gate-gap, this imbalance is diminishing with some
characteristic equilibration length $l_{eq}$. We can consider two
limits: (i) the gate-gap width $W$ is much higher than the
equilibration length, $W \gg l_{eq}$. $I-V$ curve is determined by
the equilibrium redistribution of the applied electrochemical
potential difference between ES in the gate-gap. $I-V$ trace can
be expected to be linear~\cite{alida} with the equilibrium
Buttiker slope $R_{eq}$. (ii) In the opposite case $W\ll l_{eq}$,
the charge transfer can be neglected and the applied voltage $V$
directly affects on the potential barrier between ES. $I-V$ trace
can be expected to be strongly non-linear~\cite{alida} and
asymmetric, because of the intrinsic asymmetry at the sample edge.

From our experimental results for the narrow gate-gap we can
conclude that the latter situation is realized for the integer
filling factors $\nu=3, g=2$, while for the fractional ones an
intermediate regime $W \sim l_{eq}$ takes place.

Non-linear $I-V$ curves at integer fillings $\nu=3, g=2$ can be
explained in terms of the single-particle Landau levels, bent up
by the smooth edge potential~\cite{shklovsky}, as it was reported
before~\cite{alida,relax}. The full equilibration can be achieved
only after the threshold voltage $V_{th}$ for both gate-gap
widths. It is worth to note, that we cannot expect and don't see
in the experiment $I-V$ slopes smaller than the equilibrium
$R_{eq}$, which would correspond to the additional charge transfer
between ES.

For the FQHE regime, the full equilibration at $W \sim l_{eq}$ can
be achieved at low bias $V$ (at low initial imbalances) while the
rest electrochemical imbalance at the "rejection" corner of the
gate-gap is smaller than the temperature. While increasing the
initial bias $V$, it becomes higher than the temperature,
disturbing the full equilibration and leading to the non-linear
branches. Thus, the range of the linear behavior  allows to
estimate the equilibration length for fractional fillings:
$l_{eq}\lesssim 10 \mu$m for $\nu=2/3, g=1/3$ with spin-polarized
$\nu=2/3$; $l_{eq}\sim 10 \mu$m for $\nu=2/3, g=1/3$ with
spin-unpolarized $\nu=2/3$; $l_{eq} > 10 \mu$m for $\nu=2/5,
g=1/3$. These estimations for $l_{eq}$ at 3th fractions are in good coincidence
with ones, reported before~\cite{kouwen,cunning,komiyama} for $\nu=2/3;1/3$, that
supports our analysis.

The situation for the wide gate-gap is more sophisticated. The
experimental $I-V$ traces indicate the full equilibration between
the fractional ES at $\nu=2/3$. In contrast, the $I-V$ trace is
still non-linear at $\nu=2/5$, and is situated significantly
\emph{below} the calculated line for the full equilibration, see
Fig.~\ref{IV2513}. The former could be expected for $W < l_{eq}$,
while  the latter cannot be expected for any relation between $W$
and $l_{eq}$. In terms of the picture of independent fractional ES
in the gate-gap~\cite{Beenakker}, it means that there is some
additional charge transfer between ES and they are leaving the
gate-gap region with different electrochemical potentials. It
seems to be impossible, because of the macroscopic gate-gap width
$W=800 \mu\mbox{m} \gg l_{eq} \sim 10 \mu$m. From both this fact
and the scaling between non-linear $I-V$'s with scaling
coefficient $q=2.35 \ll W_{wide}/W_{narrow}=81$ and it's
temperature behavior, we should conclude, that the model of
independent ES in the gate-gap~\cite{Beenakker} does not describe
the equilibration process at the FQHE edge at $\nu=2/5$. We should
mention here, that non-ideal contacts could not affect on the
presented result. Non-ideal contacts would lead to the non-perfect
mixing of the electrochemical potentials in
contacts~\cite{buttiker,haug}, and, thus, to rising the resistance
\emph{above} the equilibrium value.

Our experiment could be compared with the picture of interacting
fractional ES in line junction~\cite{zhulike}, where the influence
on the equilibrium conductance is expected, different for
$\nu=1/3$ and $\nu=1/5$. However, the exact
calculation~\cite{zhulike} was performed for the junction between
two quantum Hall liquids at principal filling factors
$\nu_\pm=1/(2m_\pm+1), m_\pm=0,1,...$ In our experiment, we are
working with \emph{non-principal} fractional fillings and with
equilibration at the reconstructed FQHE edge. Both points are
crucial for Ref.~\onlinecite{zhulike}, preventing us from the
direct comparison. We hope that this experiment will stimulate
further theoretical investigations.

We wish to thank  D.E.~Feldman for fruitful discussions and
A.A.~Shashkin for help during the experiment. We gratefully
acknowledge financial support by the RFBR, RAS, the Programme "The
State Support of Leading Scientific Schools", Deutsche
Forschungsgemeinschaft, and SPP "Quantum Hall Systems", under
grant LO 705/1-2. E.V.D. acknowledges support by MK-4232.2006.2
and Russian Science Support Foundation.


\begin{thebibliography}{99}
\bibitem{halperin} B. I. Halperin, Phys.\  Rev.\ B  {\bf
25}, 2185 (1982).
\bibitem{buttiker} M. B\"uttiker, Phys. Rev. B {\bf 38}, 9375 (1988).
\bibitem{haug}   R.J. Haug, Semicond. Sci. Technol.
{\bf 8}, 131 (1993).
\bibitem{alida} A. W\"urtz, R. Wildfeuer, A. Lorke,
E. V. Deviatov, and V. T. Dolgopolov, Phys. Rev. B {\bf 65},
075303 (2002); E. V. Deviatov, V. T. Dolgopolov, A. Wurtz, JETP
Lett. {\bf 79},  618 (2004).
\bibitem{laughlin} R. B. Laughlin, Phys. Rev. Lett. {\bf 50}, 1395 (1983).
\bibitem{haldane} F. D. M. Haldane, Phys. Rev. Lett. \textbf{51}, 605 (1983);
 B. I. Halperin, Phys. Rev. Lett. \textbf{52}, 1583 (1984).
\bibitem{obzor} For a review on the FQHE, see T. Chakraborty, Adv. Phys. \textbf{49}, 959
(2000).
\bibitem{macdonald} A. H. MacDonald, Phys. Rev. Lett. \textbf{64}, 220 (1990).
\bibitem{wenPRB41} Xiao-Gang Wen, Phys. Rev. B \textbf{41}, 12838 (1990).
\bibitem{chamon} C. d. C. Chamon and X. G. Wen, Phys. Rev. B \textbf{49},
8227 (1994).
\bibitem{Beenakker} C. W. J. Beenakker, Phys. Rev. Lett. \textbf{64}, 216 (1990).
\bibitem{shklovsky} D. B. Chklovskii, B. I. Shklovskii, and L. I.
Glazman, Phys. Rev. B {\bf 46}, 4026 (1992).
\bibitem{crossgates} D. A. Syphers and P. J. Stiles, Phys. Rev. B \textbf{32}, 6620 (1985);
 R. J. Haug, A. H. MacDonald, P. Streda and K. von Klitzing, Phys. Rev. Lett. \textbf{61}, 2797 (1988);
 S. Washburn, A. B. Fowler, H. Schmid and O. Kem, Phys. Rev. Lett. \textbf{61}, 2801 (1988);
  S. Komijama, H. Hira, S. Sasoand and S. Yiyamizu, Phys. Rev. B \textbf{40}, 5176 (1989).
\bibitem{milliken} F. P. Milliken, C. P. Umbach and R. A. Webb; Solid State Comm. \textbf{97}, 309
(1995).
\bibitem{pellegr} S. Roddaro, V.
Pellegrini, F. Beltram, G. Biasiol, and L. Sorba, Phys. Rev. Lett.
\textbf{93}, 046801 (2004); S. Roddaro, V. Pellegrini, F. Beltram,
L. N. Pfeiffer, and K. W. West Phys. Rev. Lett. \textbf{95},
156804 (2005).
\bibitem{chang} A. M. Chang, L. N. Pfeiffer, and K. W. West; Phys. Rev. Lett. \textbf{77}, 2538
(1996).
\bibitem{grayson}M. Grayson, D. C. Tsui, L. N. Pfeiffer, K. W. West, and A. M.
Chang, Phys. Rev. Lett. \textbf{80}, 1062 (1998); M. Hilke, D. C.
Tsui, M. Grayson, L. N. Pfeiffer, and K. W. West, Phys. Rev. Lett.
\textbf{87}, 186806 (2001).
\bibitem{fracbutt} E. V. Deviatov, V. T. Dolgopolov, A.~Lorke,
W.~Wegscheider, A.D.~Wieck, JETP Lett., \textbf{82}, 539 (2005).
\bibitem{zhulike} U. Zulicke and E. Shimshoni,Phys. Rev. B \textbf{69}, 085307
(2004); U. Zulicke and E. Shimshoni, Phys. Rev. Lett. \textbf{90},
026802 (2003).
\bibitem{ponomarenko} V. V. Ponomarenko and D. V. Averin, Phys. Rev. B \textbf{70}, 195316
(2004).
\bibitem{kouwen} L. P. Kouwenhoven, B. J. van Wees, N. C. van der Vaart, C. J. Harmans, C. E. Timmering, and C. T.
Foxon, Phys. Rev. Lett. 64, 685 (1990).
\bibitem{cunning} A. M. Chang and J. E. Cunningham, Phys. Rev. Lett. 69, 2114 (1992).
\bibitem{komiyama} T. Machida, S. Ishizuka, T. Yamazaki, S. Komiyama, K. Muraki, and Y. Hirayama,
Phys. Rev. B 65, 233304 (2002).
\bibitem{weiss} G. M\"uller, E. Diessel, D. Wiess, K. von Klitzing
K. Ploog, H. Nickel, W. Schlapp, and R. L\"osch, Surf.  Sci. {\bf
263}, 280 (1992)
\bibitem{relax} E. V. Deviatov, A. Wurtz, A. Lorke, M. Yu. Melnikov, V. T. Dolgopolov, D. Reuter, A. D.
 Wieck, Phys. Rev. B \textbf{69}, 115330 (2004).




\end{thebibliography}
\end{document}